\documentclass[journal]{IEEEtran}
\usepackage{amsmath,amssymb}
\usepackage[dvips]{graphicx}
\usepackage{amsfonts}
\usepackage[mathscr]{eucal}
\usepackage{subfigure}
\usepackage{float}
\usepackage{latexsym}
\usepackage{amsthm}
\usepackage{exscale}
\usepackage[mathscr]{eucal}
\usepackage{bm}
\usepackage[dvipsnames]{color}
\usepackage{cases}
\usepackage{color}
\usepackage{epsfig}
\usepackage[ruled]{algorithm2e}
\usepackage[verbose,nospace,sort]{cite}
\usepackage{tabularx,ragged2e}
\usepackage{multirow}
\usepackage{multicol}
\usepackage{balance}
\usepackage{caption}
\usepackage{setspace}
\usepackage{multirow}
\usepackage{multicol,afterpage,wrapfig}
\usepackage{pifont}
\usepackage{array}

\usepackage{algpseudocode}

\graphicspath{{./figs/}}

\newtheorem{lemma}{Lemma}

\newtheorem{Prop}{Proposition}

\newenvironment{myproof} {{\it{Proof:}}}{\hfill$\square$}

\begin{document}

\title{Joint Power Control and Passive Beamforming in IRS-Assisted Spectrum Sharing}
\vspace{-0cm}
\author{Xinrong Guan, \IEEEmembership{Member, IEEE,}
Qingqing Wu, \IEEEmembership{Member, IEEE,}
and Rui Zhang, \IEEEmembership{Fellow, IEEE}
\thanks{X. Guan is with the College of Communications Engineering, Nanjing, 210007, China (e-mail: geniusg2017@gmail.com). Q. Wu and R. Zhang are with the Department of Electrical and Computer Engineering, National University of Singapore, 117583, Singapore (e-mails: {elewuqq, elezhang}@nus.edu.sg). }
}

\maketitle

\begin{abstract} In cognitive radio (CR) communication systems, it is challenging to achieve high rate for the secondary user (SU) in the presence of strong cross-link interference with the primary user (PU). In this letter, we exploit the recently proposed intelligent reflecting surface (IRS) to tackle this problem. Specifically, we investigate an IRS-assisted CR system where an IRS is deployed to assist in the spectrum sharing between a PU link and an SU link. We aim to maximize the achievable SU rate subject to a given signal-to-interference-plus-noise ratio (SINR) target for the PU link, by jointly optimizing the SU transmit power and IRS reflect beamforming. Although the formulated problem is difficult to solve due to its non-convexity and coupled variables, we propose an efficient algorithm based on alternating optimization and successive convex approximation techniques to solve it sub-optimally, as well as some heuristic designs of lower complexity. Simulation results show that IRS is able to significantly improve the SU rate, even for the scenarios deemed highly challenging in conventional CR systems without the IRS.
\end{abstract}
\vspace{-0mm}

\IEEEpeerreviewmaketitle
\vspace{-3mm}
\section{Introduction}
\vspace{-0mm}
Due to the ever increasing demand for higher data rates and tremendous growth in the number of communication devices, a variety of wireless technologies have been proposed, such as massive multiple-input multiple-output (MIMO), millimeter wave (mmWave) communications, etc. However, to implement such technologies in practical systems, the increased network energy consumption and hardware cost become critical issues.
Recently, intelligent reflecting surface (IRS) has emerged as a promising technology to achieve high spectrum and energy efficiency for wireless communication cost-effectively  \cite{QQ,huang_2,QQ_1,huang2019holographic,8930608}. Specifically, IRS is a uniform planar array consisting of a large number of passive reflecting elements. By adaptively adjusting the reflection coefficient of each element, the reflected signal by IRS can be added constructively/destructively with those via other propagation paths to enhance/suppress the received signal in desired directions. 
As a result, IRS passive beamforming has been incorporated into various wireless systems such as orthogonal frequency division multiplexing (OFDM) \cite{zhang2019capacity,yifei,li2019irs}, simultaneous wireless information and power transfer (SWIPT) \cite{QQ_3,QQ_2,pan2019intelligent} and secrecy communication\cite{guan2020intelligent,Yu,Guangchi,Shen}, non-orthogonal multiple access (NOMA)\cite{zheng2020intelligent,ding1907simple,yang2019intelligent,mu2019exploiting}, etc. In particular, the fundamental squared power gain of IRS was firstly unveiled in \cite{QQ_1} and it was also shown that an interference-free zone can be established in the proximity of the IRS, by exploiting its spatial interference nulling/cancellation capability. Furthermore, it was shown in \cite{guan2020intelligent} that the artificial noise can be leveraged to improve the secrecy rate in the IRS-assisted secrecy communication, especially in presence of multiple eavesdroppers. 

\begin{figure*}[t]
	\centering
	\subfigure[IRS near the PR and the SR]{
		\includegraphics[width=1.7in]{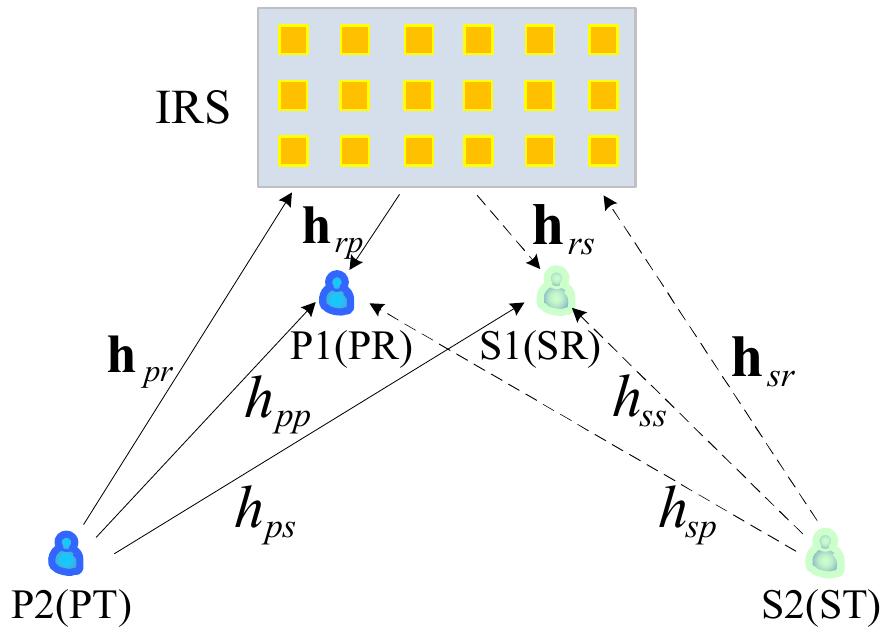}
		\label{setup_a}
	}\hspace{-3mm}
	\subfigure[IRS near the PT and the ST]{
		\includegraphics[width=1.7in]{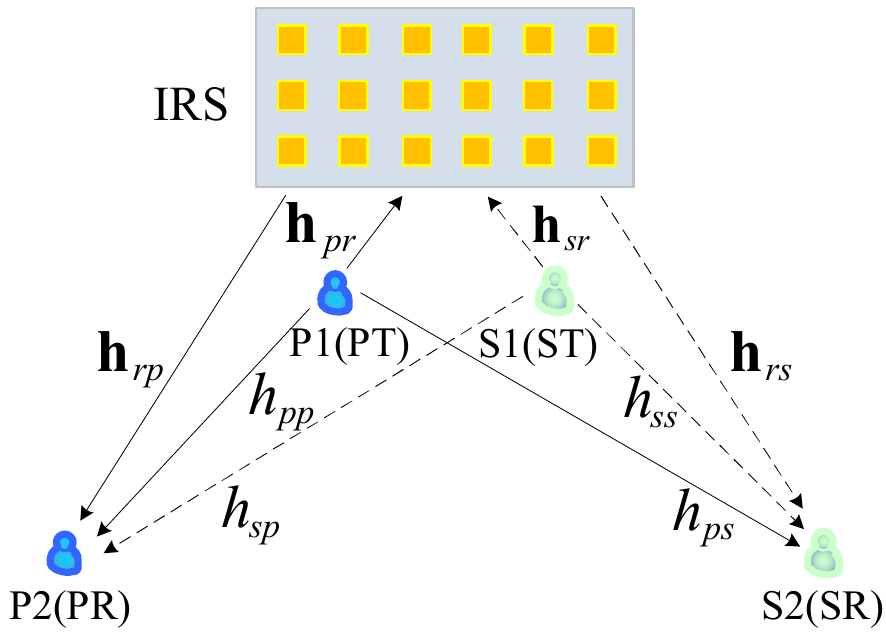}
		\label{setup_b}
	}\hspace{-2mm}
	\subfigure[IRS near the PT and the SR]{
		\includegraphics[width=1.7in]{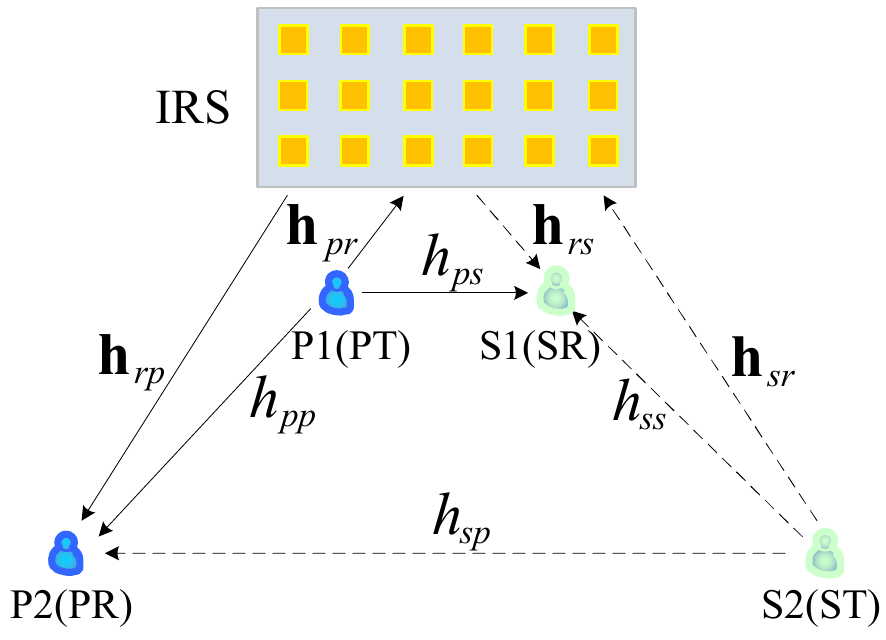}
		\label{setup_c}
	}\hspace{-2mm}
	\subfigure[IRS near the PR and the ST]{
		\includegraphics[width=1.7in]{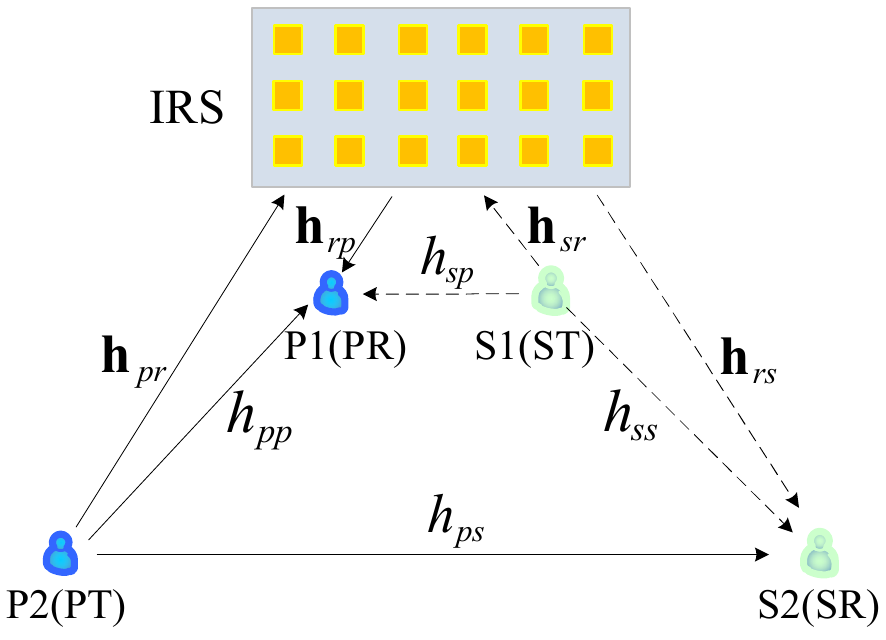}
		\label{setup_d}
	}
	\vspace{-2mm}
	\caption{IRS-assisted CR communication system subjected to (a)/(b) symmetric and (c)/(d) asymmetric interference.}
	\label{system_model}
	\vspace{0mm}
\end{figure*}

On the other hand, cognitive radio (CR) has been thoroughly investigated in the literature to solve the spectrum scarcity problem in wireless communication \cite{zhang2010dynamic}. By allowing secondary users (SUs) to share the spectrum with primary users (PUs) provided their quality of service (QoS) is ensured, the spectrum efficiency of their co-existing system can be significantly improved. However, in scenarios when the SU is located nearby the PU, e.g.,  they are in the same hotspot, the achievable SU rate becomes very limited due to the strong cross-link interference with the PU. For instance, in a CR communication system shown in Fig. \ref{system_model} (ignoring the IRS for the time being), the PU named P1 and the SU named S1 are located near each other, intending to communicate with their parters, i.e., the PU named P2 and the SU named S2, respectively. In practice, P1 and P2 can be, e.g., a mobile terminal and its serving access point/base station, whereas S1 and S2 may be a device-to-device (D2D) communication pair or Internet of Things (IoT) devices aiming to share the spectrum of the PU link for opportunistic communication. Considering each of the PU and SU nodes as a transmitter or receiver, four different communication  scenarios are shown in Fig. \ref{system_model}. In Fig. \ref{system_model}(a) and Fig. \ref{system_model}(b), the cross interfering links are symmetric, i.e. the interference from the primary transmitter (PT) to the secondary receiver (SR) is comparable to that from the secondary transmitter (ST) to the primary receiver (PR). While in the asymmetric interference scenarios Fig. \ref{system_model}(c) and Fig. \ref{system_model}(d), the interference in one of the two cross links is much stronger than that in the other. Apparently, the asymmetric interference scenario is more challenging for SU to achieve high data rate. An effective approach for the scenario Fig. \ref{system_model}(c) is to apply the successive interference cancellation (SIC) at the SR so that the strong primary interference  can be first decoded and then removed from the received signal. However, this method requires the knowledge of the codebook used by the PU and cannot deal with the most challenging scenario Fig. \ref{system_model}(d) where the strong interference from the ST to the PR is the performance bottleneck since the PR is not supposed to apply SIC in typical CR setups \cite{zhang2010dynamic}.   

By exploiting its passive beamforming for signal enhancement as well as interference suppression, IRS is potentially a promising solution to tackle the above CR challenges. This thus motivates the current work to investigate the joint ST power control and IRS reflection optimization in an IRS-assisted CR communication system shown in Fig. 1. We aim to maximize the achievable SU rate of the considered system subject to a target QoS for the PU, and  investigate how IRS helps resolve the aforementioned strong interference issue, which, has not been addressed in the literature yet. 


\vspace{-0mm}
\section{System Model and Problem Formulation}\vspace{-0mm}
\subsection{System Model}\vspace{-0mm}
As shown in Fig. 1, we consider an IRS-assisted CR/spectrum sharing system, where an SU link consisting of S1 and S2 coexists with a PU link consisting of P1 and P2, and an IRS is deployed nearby P1 and S1 to enable a communication hotspot. Assume that all nodes are equipped with a single antenna, while the number of reflecting elements at the IRS is denoted by $N$. The baseband equivalent channels from the PT(ST) to the PR, the SR and the IRS are denoted by $h_{pp}$, $h_{ps}$ and ${{\mathbf{h}}^H_{pr}} \in{\mathbb{C}^{1 \times N}}$ ($h_{sp}$, $h_{ss}$ and ${{\mathbf{h}}^H_{sr}} \in{\mathbb{C}^{1 \times N}}$),  respectively, while those from the IRS to the PR and the SR are denoted by ${{\mathbf{h}}^H_{rp}} \in{\mathbb{C}^{1 \times N}}$ and ${{\mathbf{h}}^H_{rs}} \in{\mathbb{C}^{1 \times N}}$, respectively. Let ${\mathbf{\Phi }} = \text{diag}\left( v_1,v_2,....,v_N \right)$ represent the diagonal phase-shifting matrix of the IRS \cite{QQ}, where $v_n={e^{j{\theta _n}}}$ and $\theta_n \!\in\! [0,2\pi)$ is the phase shift on the combined incident signal by its $n$-th element, $n\!\!=\!\!1,...,N$. The composite PT/ST-IRS-PR/SR channel is then modeled as a concatenation of three components, namely, the PT/ST-IRS link, the IRS reflecting with phase shifts, and the IRS-PR/SR link. The quasi-static flat-fading model is assumed for all channels. Based on the various channel acquisition methods discussed in \cite{QQ}, we assume that the channel state information (CSI) of all channels involved is perfectly known at the ST/IRS for the joint power control and passive beamforming design. Though the assumption of perfect CSI knowledge is ideal, the result obtained in this letter provides useful insights and performance bounds for practical systems with partial/imperfect CSI, which need further investigation in future work. Note that it is reasonable to assume that the CSI of P1-P2 link is known since PUs are motivated to share the CSI with the SUs/IRS for guaranteeing the QoS of the primary transmission.

Assume that the transmitted signals from the PT and the ST are given by $x_p \!\sim\! {\mathcal{CN}} \left( {0,1} \right)$ and $x_s \!\sim\! {\mathcal{CN}} \left( {0,1} \right)$, respectively. The transmit power of the PT is fixed as $p_p\!=\!P_0$, whereas that of the ST can be varied, subject to a maximum power budget denoted by $p_s \!\le\! P_{\max}$. As a result, the received signals at the PR and the SR are respectively given by\vspace{-1mm}
\begin{equation}
\small
{y_p} \!=\! {\sqrt {{p_p}} \left( {{h_{pp}}{\rm{ + }}{\bf{h}}_{rp}^H{\bf{\Phi h}}_{pr}^*} \right){x_p}}\! +\!{\sqrt {{p_s}} \left( {{h_{sp}}{\rm{ + }}{\bf{h}}_{rp}^H{\bf{\Phi h}}_{sr}^*} \right){x_s}} \! +\! {n_p},
\vspace{-0mm}
\end{equation}
\begin{equation}
\small
{y_s} \!=\!  {\sqrt {{p_s}} \left( {{h_{ss}}{\rm{ + }}{\bf{h}}_{rs}^H{\bf{\Phi h}}_{sr}^*} \right){x_s}}  \!+\!{\sqrt {{p_p}} \left( {{h_{ps}}{\rm{ + }}{\bf{h}}_{rs}^H{\bf{\Phi h}}_{pr}^*} \right){x_p}}  \!+\! {n_s},
\vspace{-1mm}
\end{equation}
where ${n_p}\!\sim\! {\mathcal{CN}} \left( {0,\sigma_p^2} \right)$ and ${n_s}\!\sim\! {\mathcal{CN}} \left( {0,\sigma_s^2} \right)$ denote the complex additive white Gaussian noise (AWGN) at the PR and the SR, respectively. Denoting ${{\mathbf{h}}_{irj}} \!=\! \text{diag}\left( {{\mathbf{h}}_{rj}^H} \right){{\mathbf{h}}_{ir}^*}$ and ${{{\mathbf{ v}}}^H}\!=\!\left[v_1, v_2,...,v_N \right]$, we have ${{\mathbf{h}}_{rj}^H{\mathbf{\Phi h}}_{ir}^* = {{\mathbf{v}}^H}{{\mathbf{h}}_{irj}}}$, $i,j \in \{p,s\}$. Thus, the signal-to-interference-plus-noise ratio (SINR) at the PR and the SR can be respectively expressed as\vspace{-1mm}
\begin{equation}
\small
\gamma_p=\frac{{{p_p}{{\left| {{{\bf{v}}^H}{{\bf{h}}_{prp}} + {h_{pp}}} \right|}^2}}}{{{p_s}{{\left| {{{\bf{v}}^H}{{\bf{h}}_{srp}} + {h_{sp}}} \right|}^2} + \sigma _p^2}},
\vspace{-1mm}
\end{equation}
\begin{equation}\vspace{-0mm}
\small
\gamma_s=\frac{{{p_s}{{\left| {{{\bf{v}}^H}{{\bf{h}}_{srs}} + {h_{ss}}} \right|}^2}}}{{{p_p}{{\left| {{{\bf{v}}^H}{{\bf{h}}_{prs}} + {h_{ps}}} \right|}^2} + \sigma_s^2}}.
\end{equation}
To guarantee the QoS of the PU link, we impose the SINR constraint at PR as $\gamma_p \!\ge\! \gamma _\text{th}$, , where $\gamma _\text{th}$ is the PU SINR target. 

\vspace{-4mm}
\subsection{Problem Formulation}\vspace{-0mm}
We aim to maximize the achievable SU rate $R_s\!=\!\rm{log_2}(1+\gamma_s)$ in bits/second/Hertz (bps/Hz) via joint the power control at the ST with the reflect beamforming at the IRS, subject to the SINR constraint at the PR. Thus, the optimization problem is formulated as

~\\\vspace{-8mm}
{\small
	\begin{align*}	
	\left( {{\rm{P0}}} \right):\mathop {\max }\limits_{{p_s},{\bf{v}}}  ~~& {\rm{log_2}}\left(1+\frac{{{p_s}{{\left| {{{{\bf{ v}}}^H}{{{\bf{h}}}_{srs}}+h_{ss}} \right|}^2}}}{{{p_p}{{\left| {{{\bf{v}}^H}{{\bf{h}}_{prs}} + {h_{ps}}} \right|}^2} + \sigma_s^2}}\right)\notag\\
	\text{s.t.}~~~&\frac{{{p_p}{{\left| {{{\bf{v}}^H}{{\bf{h}}_{prp}} + {h_{pp}}} \right|}^2}}}{{{p_s}{{\left| {{{\bf{v}}^H}{{\bf{h}}_{srp}} + {h_{sp}}} \right|}^2} + \sigma _p^2}} \ge {\gamma _ \text{th}},\\
	&{p_s} \le {P_{\max}},\\
	&\left| {{{  v}_n}} \right| = 1,n = 1,..,N.
	\end{align*}
}\vspace{-7mm}\\
(P0) is difficult to solve due to the non-concave objective function as well as the coupled optimization variables. However, we observe that when one of $p_s$ and ${\bf{v}}$ is fixed, the resultant problems can be efficiently solved. This thus motivates us to propose an alternating optimization (AO) based algorithm to solve (P0) sub-optimally, by iteratively optimizing one of $p_s$ and ${\bf{v}}$ with the other being fixed at each iteration until the convergence is reached.
\vspace{-0mm}
\section{Proposed Solutions}\vspace{-0mm}
In this section, we propose the AO based algorithm first, followed by some heuristic designs for reducing the complexity and signaling overhead.\vspace{-3mm}
\subsection{Alternating Optimization Based Joint Design}\vspace{-0mm}
\subsubsection{Optimizing $p_s$ for Given $\bf{v}$}Denoting $\small{\bf{\bar h}}_{ij} \!=\! \left[ 	{{{\mathbf{h}}_{irj}^T}}~~{{{h}}_{ij}} 
\right]^T$ and ${{{\mathbf{\tilde v}}}^H} \!= \!{e^{j\varpi}}\left[ {{{\mathbf{v}}^H}~1} \right]$ where ${\varpi}$ is an arbitrary phase rotation, then the equivalent channel power gain is given by $\alpha_{ij}\!\!=\!\!{{\left| {{{{\mathbf{\tilde v}}}^H}{\bf{\bar h}}_{ij}} \right|}^2}$,  $i,j \!\in\! \{s,p\}$. Thus, (P0) can be rewritten as \vspace{-4mm}\\
\begin{small}
	\begin{align*}	
	\left( {{\rm{P1}}} \right):\mathop {\max }\limits_{{p_s}}  ~~& {\rm{log_2}}\Big(1+\frac{{{p_s}\alpha_{ss}}}{{{p_p}\alpha_{ps} + \sigma_s^2}}\Big)\notag\\
	\text{s.t.}~~~&\frac{{{p_p}\alpha_{pp}}}{{{p_s}\alpha_{sp} + \sigma _p^2}} \ge {\gamma _ \text{th}},\\
	&{p_s} \le {P_{\max}}.
	\end{align*}
\end{small}
\vspace{-6.5mm}\\
It can be observed that the objective function in (P1) is a monotonically increasing function of $p_s$, thus the optimal $p_s$ can be obtained in closed-form as\vspace{-1.5mm}
\begin{equation}\small
	\label{p_s}
	p_s^\star=\max\left(0, \min \left( {\left( {{P_0}{\alpha _{pp}}/{\gamma _\text{th}} - \sigma _p^2} \right)/{\alpha _{sp}}},{P_{\max }} \right)\right).\vspace{-1.5mm}
\end{equation}
\subsubsection{Optimizing $\bf{v}$ for Given $p_s$}For given $p_s$, denoting $ {{\bf{H}}_{ij}} = {{{\bf{\bar h}}}_{ij}}{\bf{\bar h}}_{ij}^H$, ${{\bf{A}}_{jj}} \!=\! {p_j}{{\bf{H}}_{jj}}$ and ${{\bf{B}}_{ij}}\! =\! {p_i}{{\bf{H}}_{ij}}\! +\! {{\bf{I}}_{N + 1}}\sigma _j^2/\left( {N \!+ \!1} \right)$, where $i,j \in \left\{ {p,s} \right\}$ and ${{\bf{I}}_{N }}$ is the $N \times N $ identity matrix, then (P0) can be reformulated as\\\vspace{-4mm}
\begin{small}
	\begin{align}
	\left( {{\rm{P}}2.1} \right):\mathop {\max }\limits_{{\bf{\tilde v}}} ~~&\frac{{{{{\bf{\tilde v}}}^H}{{\bf{A}}_{ss}}{\bf{\tilde v}}}}{{{{{\bf{\tilde v}}}^H}{{\bf{B}}_{ps}}{\bf{\tilde v}}}}\notag\\
	{\rm{s.t.~~~}}&\frac{{{{{\bf{\tilde v}}}^H}{{\bf{A}}_{pp}}{\bf{\tilde v}}}}{{{{{\bf{\tilde v}}}^H}{{\bf{B}}_{sp}}{\bf{\tilde v}}}} \ge {\gamma _\text{th}},\tag{6.1}\label{C3}\\
	&\left| {{{\tilde v}_n}} \right| = 1,n = 1,...,N + 1.\tag{6.2}\label{C4}
	\end{align}
\end{small}
\vspace{-6mm}\\
However, (P2.1) is still difficult to solve since it is a fractional quadratically constrained quadratic programing (QCQP) problem with unit modulus constraints. To overcome such difficulty, we apply the successive convex approximation (SCA) technique to transform (P2.1) into a series of more tractable approximated subproblems. In particular, we resort to the following lemma to derive a lower bound on the objective function and the left-hand-side (LHS) of (\ref{C3}), respectively.\vspace{-2mm}
\begin{lemma}\label{Lemma_1}
Denoting $f\!\left(\!{{\bf{A}}_{jj}},{{\bf{B}}_{ij}}, {\bf{\tilde v}} \right) \!\!=\!\! \frac{{{{{\bf{\tilde v}}}^H}{{\bf{A}}_{jj}}{\bf{\tilde v}}}}{{{{{\bf{\tilde v}}}^H}{{\bf{B}}_{ij}}{\bf{\tilde v}}}}$, $i,j \!\!\in \!\!\left\{ {p,s} \right\}, i \!\neq\! j$, then the following inequality holds for any given ${\bf{\tilde v}}_0$,\vspace{-1mm}
\setcounter{equation}{6}
\begin{equation}\vspace{-2mm}
\small
\label{lemma1}
	f\left({{\bf{A}}_{jj}},{{\bf{B}}_{ij}}, {\bf{\tilde v}} \right) \ge 2{\mathop{\rm Re}\nolimits} \big\{ {{\bf{w}}_{jj}^H{\bf{\tilde v}}} \big\} + {d_{jj}},
\end{equation}
where\vspace{-0.5mm}
{
\small
\begin{align}
	{\bf{w}}_{jj}^H &\!=\frac{{{\bf{\tilde v}}_0^H{{\bf{A}}_{jj}}}}{{{\bf{\tilde v}}_0^H{{\bf{B}}_{ij}}{{{\bf{\tilde v}}}_0}}}\! -\! {\bf{\tilde v}}_0^H\left( {{{\bf{B}}_{ij}} \!- \!{\lambda _{{{\bf{B}}_{ij}}}}{{\bf{I}}_{N + 1}}} \right)\frac{{{\bf{\tilde v}}_0^H{{\bf{A}}_{jj}}{{{\bf{\tilde v}}}_0}}}{{{{\left( {{\bf{\tilde v}}_0^H{{\bf{B}}_{ij}}{{{\bf{\tilde v}}}_0}} \right)}^2}}},\label{w}\\ 
	{d_{jj}} &= \!-\! \left[ {2{\lambda _{{{\bf{B}}_{ij}}}}\left( {N\! + \!1} \right) \!- \!{\bf{\tilde v}}_0^H{{\bf{B}}_{ij}}{{{\bf{\tilde v}}}_0}} \right]\frac{{{\bf{\tilde v}}_0^H{{\bf{A}}_{jj}}{{{\bf{\tilde v}}}_0}}}{{{{\left( {{\bf{\tilde v}}_0^H{{\bf{B}}_{ij}}{{{\bf{\tilde v}}}_0}} \right)}^2}}}\label{d},
\end{align}
}\vspace{-3mm}\\
and ${\lambda _{{{\bf{B}}}_{ij}}}$ denotes the maximum eigenvalue of ${{{\bf{B}}}_{ij}}$. 
\end{lemma}\vspace{-2mm}
\begin{myproof}
	See Appendix A.
\end{myproof}\vspace{0.5mm}\\ 
\setlength{\intextsep}{0pt}
\setlength{\textfloatsep}{-1mm} 
\begin{algorithm}[t]
	\caption{Alternating optimization for solving (P0)}
	\label{Algorithm1}
	\LinesNumbered 
	\DontPrintSemicolon	
	\SetAlgoHangIndent{0.1em}
	\KwIn{$N$, $P_{\rm max}$,  $\sigma_j^2$, $\gamma_\text{th}$, ${\bf{h}}_{irj}$, ${{h}}_{ij}$, $i,j \in \{p,s\}$.}
	\KwOut{$p_s$, $\bf{v}$.}	
	Initialize the phase-shift  vector as ${{{\bm{\theta}}}^{(0)}}\!=\![\theta_1,...,\theta_N]^T$.\\
	Set $l=1$, ${{{\mathbf{v}}}^{(0)}} = \big[e^{j\theta_1},...,e^{j\theta_N}\big]^T$ and ${{{\mathbf{\tilde v}}}^{(0)}} = \Big[ {\begin{matrix}
		{{\mathbf{v}}^{(0)}} \\ 
		1 
		\end{matrix}} \Big]$.\vspace{-1mm}\\ 
	\Repeat{\rm the objective value in (P0) reaches convergence.}
	{
		For given ${\bf{\tilde v}}^{(l-1)}$, obtain $p_s^{(l)}$ according to (\ref{p_s}).
		\\
		{\spaceskip=0.17em\relax For given $p_s^{(l)}$, update ${\bf{A}}_{jj}$ and ${\bf{B}}_{ij}$. Set ${\bf{\tilde v}}_0={\bf{\tilde v}}^{(l-1)}$ and update ${\bf{w}}_{ss}^H$, ${\bf{w}}_{pp}^H$, $d_{ss}$ and $d_{pp}$ according to (\ref{w}) and (\ref{d}).}\\ 
		\Repeat{\rm{\spaceskip=0.2em\relax the objective value in (P2.3) reaches convergence.}}
		{
			If $\gamma_p(0)\ge\gamma_{\rm{th}}$, $u_n^\star\!=\!e^{j\angle\left( {{w_{ss}}(n)} \right)}$. Otherwise, solve $\lambda^\star$ via bisection search and obtain ${\bf{u}}^\star$ in (\ref{u_n}). \\
			Update ${\bf{\tilde v}}_0$ as ${\bf{\tilde v}}_0={\bf{u}}^\star$ and ${\bf{w}}_{ss}^H$, ${\bf{w}}_{pp}^H$, $d_{ss}$ and $d_{pp}$ according to (\ref{w}) and (\ref{d}).
		}	        
		Update ${\bf{\tilde v}}^{(l)}={\bf{u}}^\star$ and $l=l+1$.
	}	        
	Obtain ${\bf{v}}$ from ${\bf{\tilde v}}$ according to (\ref{v}).\vspace{-1mm}
\end{algorithm}
According to {\bf{Lemma \ref{Lemma_1}}}, we have $\frac{{{{{\bf{\tilde v}}}^H}{{\bf{A}}_{ss}}{\bf{\tilde v}}}}{{{{{\bf{\tilde v}}}^H}{{\bf{B}}_{ps}}{\bf{\tilde v}}}} \!\ge\! 2{\mathop{\rm Re}\nolimits} \big\{ {{\bf{w}}_{ss}^H{\bf{\tilde v}}} \big\} \!+\! {d_{ss}}$
and $\frac{{{{{\bf{\tilde v}}}^H}{{\bf{A}}_{pp}}{\bf{\tilde v}}}}{{{{{\bf{\tilde v}}}^H}{{\bf{B}}_{sp}}{\bf{\tilde v}}}} \!\ge\! 2{\mathop{\rm Re}\nolimits} \left\{ {{\bf{w}}_{pp}^H{\bf{\tilde v}}} \right\} + {d_{pp}}$. Then, we consider the following optimization problem for given ${\bf{\tilde v}}_0$,\vspace{-1.5mm}
\begingroup
\addtolength{\jot}{-0.5mm}
\begin{equation*}
	\small
	\begin{aligned}
	\left( {{\rm{P}}2.2} \right):\mathop {\max }\limits_{{\bf{\tilde v}}} ~~&{{\rm{2}}{\mathop{\rm Re}\nolimits} \big\{ {{\bf{w}}_{ss}^H{\bf{\tilde v}}} \big\} + {d_{ss}}}\notag\\
	{\rm{s.t.~~~}}&{\rm{2}}{\mathop{\rm Re}\nolimits} \big\{ {{\bf{w}}_{pp}^H{\bf{\tilde v}}} \big\} + {d_{pp}} \ge {\gamma _\text{th}},\\
	&\left| {{{\tilde v}_n}} \right| = 1,n = 1,...,N + 1.
	\end{aligned}	\vspace{-1.5mm}
\end{equation*}
\endgroup
The remaining difficulty in solving (P2.2) lies in the non-convexity of the unit modulus constraints. In the following, we first construct a convex problem (P2.3) which is obtained by properly relaxing the non-convex constraints in (P2.2), then show that the optimal solution to (P2.3) must satisfy the unit modulus constraints and thus is also optimal to (P2.2).\vspace{-1mm}	

{\vspace{-4mm}
\small
\begingroup
\addtolength{\jot}{-0.5mm}
	\begin{align}
	\left( {{\rm{P}}2.3} \right):\mathop {\max }\limits_{{\bf{u}}} ~~&{{\rm{2}}{\mathop{\rm Re}\nolimits} \big\{ {{\bf{w}}_{ss}^H{\bf{u}}} \big\} + {d_{ss}}}\notag\\
	{\rm{s.t.~~~}}&{\rm{2}}{\mathop{\rm Re}\nolimits} \big\{ {{\bf{w}}_{pp}^H{\bf{u}}} \big\} + {d_{pp}} \ge {\gamma _\text{th}},\tag{10.1}\label{C7}\\
	&\left| {{{u}_n}} \right|^2 \le 1,n = 1,...,N + 1.\label{C8}\tag{10.2}
	\end{align}\setcounter{equation}{10}\vspace{-6mm}
\endgroup
}\\
Denoting the dual variables associated with constraints (\ref{C7}) and (\ref{C8}) by $\lambda \!\!\ge\!\! 0$ and $\mu_n \!\ge\! 0$, $n\!\!=\!\!1,\!...\!,N\!+\!1$, respectively, and the $n$-th element in ${\bf{w}}_{ss}$ and ${\bf{w}}_{pp}$ by $\!w_{ss}(n)\!$ and $\!w_{pp}(n)\!$, respectively. A closed-form expression for the optimal solution to (P2.3) is given by the following proposition. \vspace{-1mm}
\begin{Prop}
The optimal solution to (P2.3) is ${u_n^\star}\! =\! {e^{j\angle \left( {{w_{ss}}(n) + \lambda {w_{pp}}(n)} \right)}}$, where $\angle(x)$ denotes the phase of $x$.
\end{Prop}\vspace{-2.5mm}
\begin{myproof}
	See Appendix B.
\end{myproof}\vspace{0.5mm}\\
Based on {\bf{Proposition 1}}, we have ${u_n^\star}\! =\! {e^{j\angle \left( {{w_{ss}}(n)} \right)}}$, $\forall \lambda\ge0$, if $\angle \left( {{w_{ss}}(n)} \right)=\angle \left( { {w_{pp}}(n)} \right)$.
Otherwise, we consider the following two cases to derive the optimal $\lambda$, denoted by $\lambda^\star$. Rewrite the LHS of (\ref{C7}) as \vspace{-1.5mm}
\begin{equation}\small\vspace{-1mm}
\label{gamma_p_lambda}
{\gamma _p}\left( \lambda  \right)\! =\! 2\!\sum\nolimits_n\! {{\mathop{\rm Re}\nolimits} \!\left\{\! {w_{pp}^*(n){e^{j\angle \left( {{w_{ss}}(n) +\lambda {w_{pp}}(n)} \!\right)\!}}} \!\right\}}\! \! +\! {d_{pp}}.
\end{equation}

Case (1): Assume that $\lambda^\star\!=\!0$, then  $u_n^\star$ is rewritten as $u_n^\star\!=\!e^{j\angle\left( {{w_{ss}}(n)} \right)}$, which has to satisfy the SINR constraint in (\ref{C7}), i.e. $\gamma_p(0)\ge\gamma_{\rm{th}}$ must hold. Otherwise, it should be Case (2).

Case (2): $\lambda^\star>0$, then it follows that the equality ${\gamma _p}\left( \lambda^\star  \right)={\gamma _\text{th}}$ must hold according to the complementary slackness condition. To find  $\lambda^\star$ satisfying ${\gamma _p}\left( \lambda^\star  \right)={\gamma _\text{th}}$, we provide the following lemma.\vspace{-1.5mm}
\begin{lemma}\label{Lemma_2}
	$\!{\gamma _p}\left( \lambda\right)\!$ is a monotonically increasing function of $\lambda$ if $\angle \left( {{w_{ss}}(n)} \right) \ne \angle \left( { {w_{pp}}(n)} \right)$.
\end{lemma}\vspace{-1mm}
\begin{myproof}
See Appendix C.
\end{myproof}\vspace{1mm}\\ 
Based on {\bf{Lemma \ref{Lemma_2}}}, if $\gamma_p(\infty)\!<\!\gamma_{\rm{th}}$, ${\gamma _p}\left( \lambda^\star  \right)\!=\!{\gamma _\text{th}}$ is infeasible and so is problem (P2.3). Otherwise, $\lambda^\star$ can be obtained by using the bisection search method. After $\lambda^\star$ is obtained, ${u_n^\star}$ in Case (2) is given by 
\vspace{-1.5mm}
\begin{equation}\small
\label{u_n}
{u_n^\star} = {e^{j\angle \left( {{w_{ss}}(n) + \lambda^\star {w_{pp}}(n)} \right)}}, n = 1,...,N+1.\vspace{-2.5mm}
\end{equation}

Note that ${\bf{u}}^\star$ is optimal to (P2.3) and satisfies all the constraints in (P2.2), thus ${\bf{\tilde v}}^\star\!=\!{\bf{u}}^\star$ is the optimal solution to (P2.2). Otherwise, it implies that there exists a feasible solution to (P2.2) (also feasible to (P2.3)) that achieves a larger objective value than the optimal objective value of (P2.3) achieved by ${\bf{u}}^\star$, which causes contradiction. As such, an approximate solution to (P2.1) is obtained by successively solving (P2.3) and updating ${\bf{\tilde v}}_0$ as ${\bf{\tilde v}}_0={\bf{ u}}^\star$. Finally, the reflection coefficients are obtained as
\vspace{-1mm}
\begin{equation}\small
\label{v}
{{{v}}_n^\star} =e^{j\, \angle ({{{{{{\tilde v^\star}}}_n}}}/{{{{{{\tilde v^\star}}}_{{N + 1}}}}})}, n = 1,...,N.
\vspace{-2.5mm}
\end{equation}

\subsubsection{Overall Algorithm}
To summarize, the overall iterative algorithm to solve (P0) is given in Algorithm \ref{Algorithm1}. The main complexity of Algorithm \ref{Algorithm1} is due to computing ${\bf{w}}_{ss}^H$, ${\bf{w}}_{pp}^H$, $d_{ss}$ and $d_{pp}$ in Step 5 and Step 8. Specifically, denoting the number of iterations required for the objective value of (P2.3) and (P0) to reach convergence by $L_1$ and $L_2$, respectively, the overall complexity is obtained as $\mathcal{O}(L_2(L_1+1)(N+1)^3)$.

\vspace{-5mm}
\subsection{Low Complexity Two-Stage Designs}\vspace{-1mm}
Next, low complexity two-stage designs are proposed to reduce the implementation overhead. Specifically, in the first stage, the passive beamforming is optimized to maximize the channel power gain of the equivalent PT-PR/ST-SR (desired) link, i.e. $\alpha_{pp}$ and $\alpha_{ss}$, respectively, or to minimize that of the equivalent PT-SR/ST-PR (interference) link, i.e. $\alpha_{ps}$ and $\alpha_{sp}$, respectively. In the second stage, the transmit power at the ST is optimized similarly as (\ref{p_s}). The passive beamforming vector design in the first stage is specified as follows. 
\subsubsection{Signal power maximization based designs}Assuming that ${{\bf{ v}}}$ is designed to maximize $\alpha_{pp}$, the optimization problem is formulated as\vspace{-9mm}\\

\begin{small}
	\begin{align*}
	\left( {{\rm{P3}}} \right):\mathop {\max }\limits_{{\bf{v}}}  ~~& {{{\left| {{{{\bf{ v}}}^H}{{{\bf{h}}}_{prp}}} +h_{pp}\right|}^2}}\\
	\text{s.t.}~~~&\left| {{{  v}_n}} \right| = 1,n = 1,..,N.
	\end{align*}
\end{small}
\vspace{-6mm}\\
The optimal solution is given by\cite{QQ_1}\vspace{-1mm}
\begin{equation}\small
\vspace{-1mm}
\label{v_n2}
{v_n^\star} = {e^{j \left( \angle{{h_{pp}} - \angle{{{h}}_{prp}}(n)} \right)}}, n = 1,...,N.\vspace{-1.5mm}
\end{equation}
Similarly, the optimal  ${{\bf{ v}}}$ maximizing $\alpha_{ss}$ can be obtained.
\subsubsection{Interference power minimization based designs}First, design  ${{\bf{ v}}}$ to minimize $\alpha_{sp}$, which is formulated as\vspace{-8.5mm}\\

\begin{small}
	\begin{align*}
	\left( {{\rm{P4}}} \right):\mathop {\min }\limits_{{\bf{v}}}  ~~& {{{\left| {{{{\bf{ v}}}^H}{{{\bf{h}}}_{srp}}} +h_{sp}\right|}^2}}\\
	\text{s.t.}~~~&\left| {{{  v}_n}} \right| = 1,n = 1,..,N.
	\end{align*}
\end{small}\vspace{-9.5mm}\\

Case (1): If $\sum\nolimits_n{\mathop{|{{{h}}_{srp}}(n)|}\nolimits}\le|h_{sp}|$, which implies that the maximum channel power gain of the reflecting ST-IRS-PR link is no larger than that of the direct ST-PR link, then we have $\mathop {\min }\limits_{{\bf{v}}}~ \alpha_{sp}=\left(|h_{sp}|-\sum\nolimits_n{\mathop{|{{{h}}_{srp}}(n)|}\nolimits}\right)^2$. This suggests that the interference cannot be completely canceled in this case. The optimal reflection coefficients are thus given by \vspace{-1mm}
\begin{equation}
\label{v_n}
{v_n^\star} = {e^{j \left( \pi+\angle{{h_{sp}} - \angle{{{h}}_{srp}}(n)} \right)}}, n = 1,...,N.\vspace{-1mm}
\end{equation}

Case (2): If $\sum\nolimits_n{\mathop{|{{{h}}_{srp}}(n)|}\nolimits}>|h_{sp}|$, define a matrix as ${{\bf{\tilde V}}}\!=\!{{\bf{\tilde v}}}{{\bf{\tilde v}}^H}$, then it follows that ${{\bf{\tilde V}}} \!\!\succeq\!\! 0$ and ${\rm rank}({{\bf{\tilde V}}})\!\!=\!\!1$. By applying the semidefinite relaxation (SDR) to relax the non-convex rank-1 constraint, (P4) is reduced to\vspace{-8.5mm}\\

\begin{small}
	\begin{align*}
	\left( {{\rm{P4.1}}} \right):\mathop {\min }\limits_{{\bf{\tilde V}}}  ~~& {\rm Tr}({\bf{H}}_{sp}{\bf{\tilde V}})\\
	\text{s.t.}~~~& {\bf{\tilde V}}_{n,n}   = 1,n = 1,..,N+1,
	\end{align*}
\end{small}
\vspace{-6mm}\\
where ${\bf{H}}_{sp}$ is given in Section III-A-2). (P4.1) can be efficiently solved by using a convex optimization solver. If the obtained ${{\bf{\tilde V}}}$ is of rank-1, the optimal reflection coefficients can be obtained by applying eigenvalue decomposition as ${{\bf{\tilde V}}}={{\bf{\tilde v}}}{{\bf{\tilde v}}}^H$. Otherwise, the suboptimal solution can be recovered from ${\bf{\tilde V}}$ via Gaussian randomization  \cite{QQ_1}. Similarly, the optimal/suboptimal ${{\bf{ v}}}$ minimizing $\alpha_{ps}$ can be obtained.

\vspace{-0mm}
\section{Simulation Results}\vspace{-0mm}
To study the effect of IRS on the CR communication system, we consider three different setups, namely Setup (1), (2) and (3), corresponding to the scenarios Fig. 1(a), (c) and (d), respectively. Note that the scenario Fig. 1(b) is omitted since its result is similar to that of Fig. 1(a). The deployment of P1, P2, S1, S2 and IRS is illustrated in Fig. 2. We assume that the system operates on a carrier frequency of 750 MHz with the wavelength $\lambda_c\!\!=\!\!0.4$ m and the path loss at the reference distance $d_0\!=\!$1 m is given by $L_0\!=\!-30$ dB. Suppose that the IRS is equipped with a uniform planar array with 6 rows and 10 columns, and the element spacing is $\Delta d\!=\! 3\lambda_c/8$. The noise power is set as $\sigma_p^2\!=\!\sigma_s^2\!=\!-$105 dBm. The channel between P1 and P2 is generated by ${{{h}}^H_{pp}} \!\!= \!\! ({{L_0}D_{pp}^{ - {c_{pp}}}})^{1/2} {{{g}}_{pp}}$, where $D_{pp}$ denotes the distance from P1 to P2 and $c_{pp}$ denotes the corresponding path loss exponent. The small-scale fading component ${{{g}}_{pp}}$ is assumed to be Rician fading and given by
${\small\! {{{g}}_{pp}} \!=\! \sqrt {{{{\beta _{pp}}}}/({{1 \!+\! {\beta _{pp}}}})} {{g}}_{pp}^{\rm LoS} \!+ \!\sqrt {{1}/({{1 \!+\! {\beta _{pp}}}})}{{g}}_{pp}^{\rm NLoS}\!}$, 
where $\beta_{pp}$ is the Rician factor, ${{g}}_{pp}^{\rm LoS}$ and $g_{pp}^{\rm NLoS}$ represent the deterministic line-of-sight (LoS) and Rayleigh fading/non-LoS (NLoS) components, respectively. The same channel model is adopted for all other channels in general. 
In particular, we assume that the channels among the ground nodes, i.e. P1, S1, P2 and S2, as well as those between IRS and P2/S2 have no LoS component due to  terrestrial rich scattering/Rayleigh fading with the path loss exponents and Rician factors set as 3 and 0, respectively,  whereas the channels between IRS and P1/S1 are LoS due to the higher altitude of the IRS and its short distances with P1/S1, with the path loss exponents and Rician factors set as 2 and $\infty$, respectively.
\begin{figure}[t]
	\centering
	\includegraphics[width=2.45in]{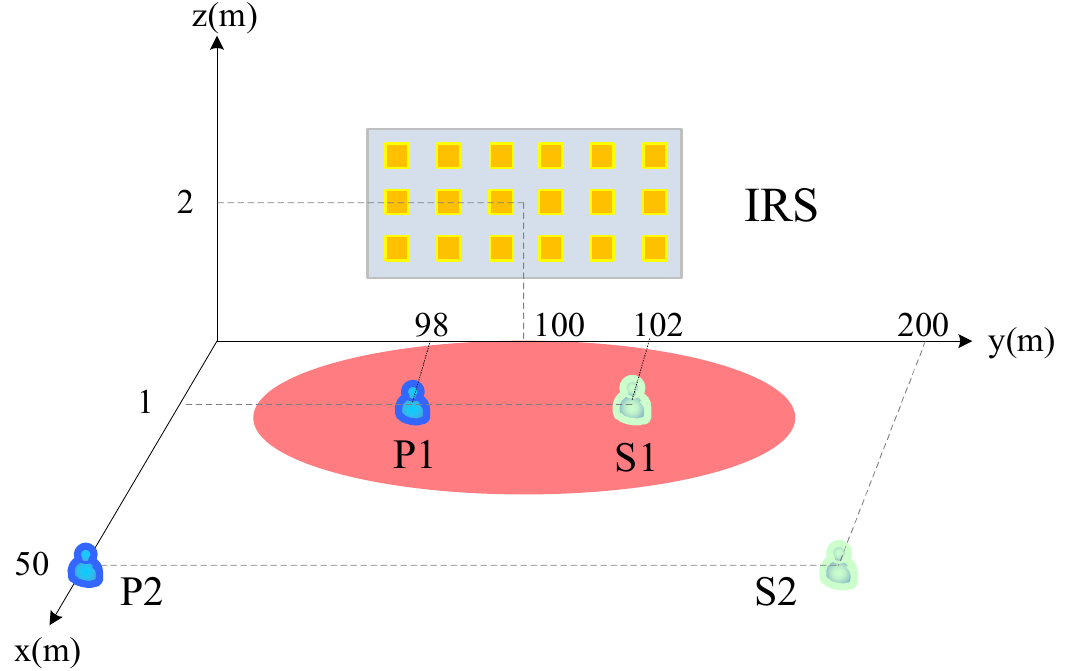}
	\label{simulation_setup}\vspace{-1mm}
	\caption{Simulation setup.}\vspace{1mm}
\end{figure}

Besides the proposed AO based design (IRS, AO) and four low-complexity designs (IRS, max $\alpha_{pp}$), (IRS, max $\alpha_{ss}$), (IRS, min $\alpha_{sp}$)  and (IRS, min $\alpha_{ps}$), the cases without IRS and with/without SIC at the SR, i.e. (no-IRS, w/ SIC) and (no-IRS, w/o SIC), are also considered for performance comparison and showing the benefit of using IRS.  

\begin{figure*}[t]
	\centering
	\subfigure[Setup (1): corresponding to Fig. 1(a)]{
		\includegraphics[width=2.1in]{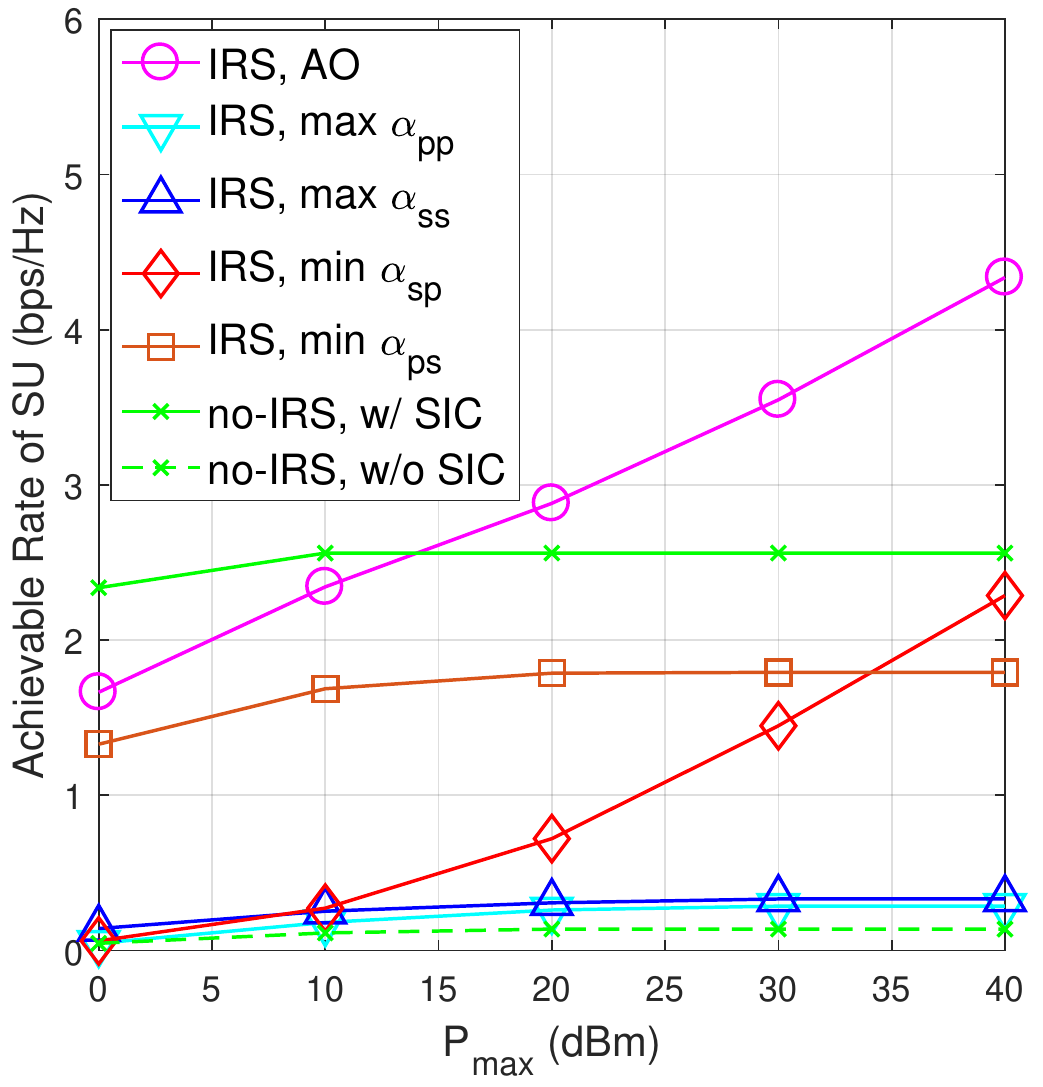}
		\label{fig:rate_setup_a_2}
	}	\hspace{0.5mm}
	\subfigure[Setup (2): corresponding to Fig. 1(c)]{
		\includegraphics[width=2.1in]{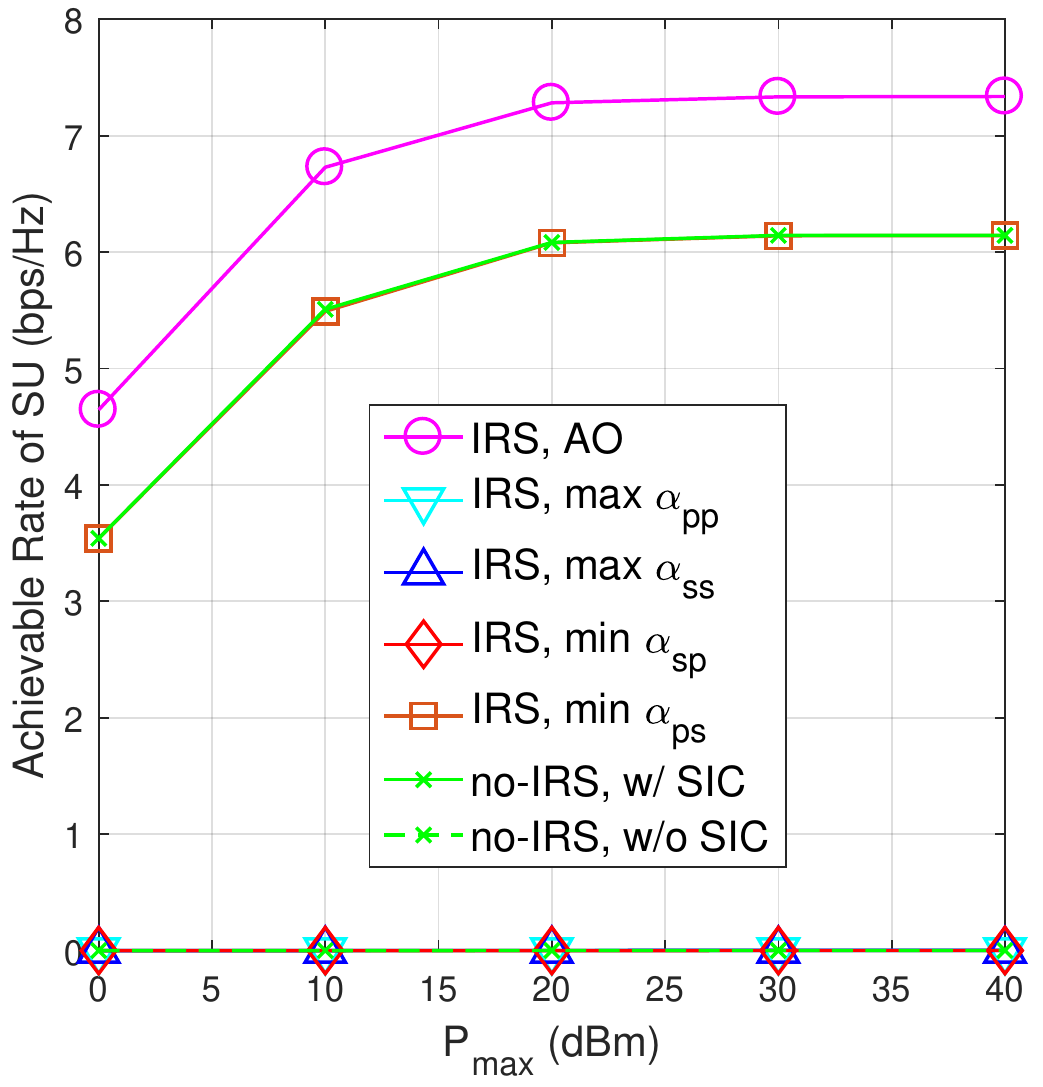}
		\label{fig:rate_setup_b_2}
	}\hspace{1.5mm}
	\subfigure[Setup (3): corresponding to Fig. 1(d)]{
		\includegraphics[width=2.1in]{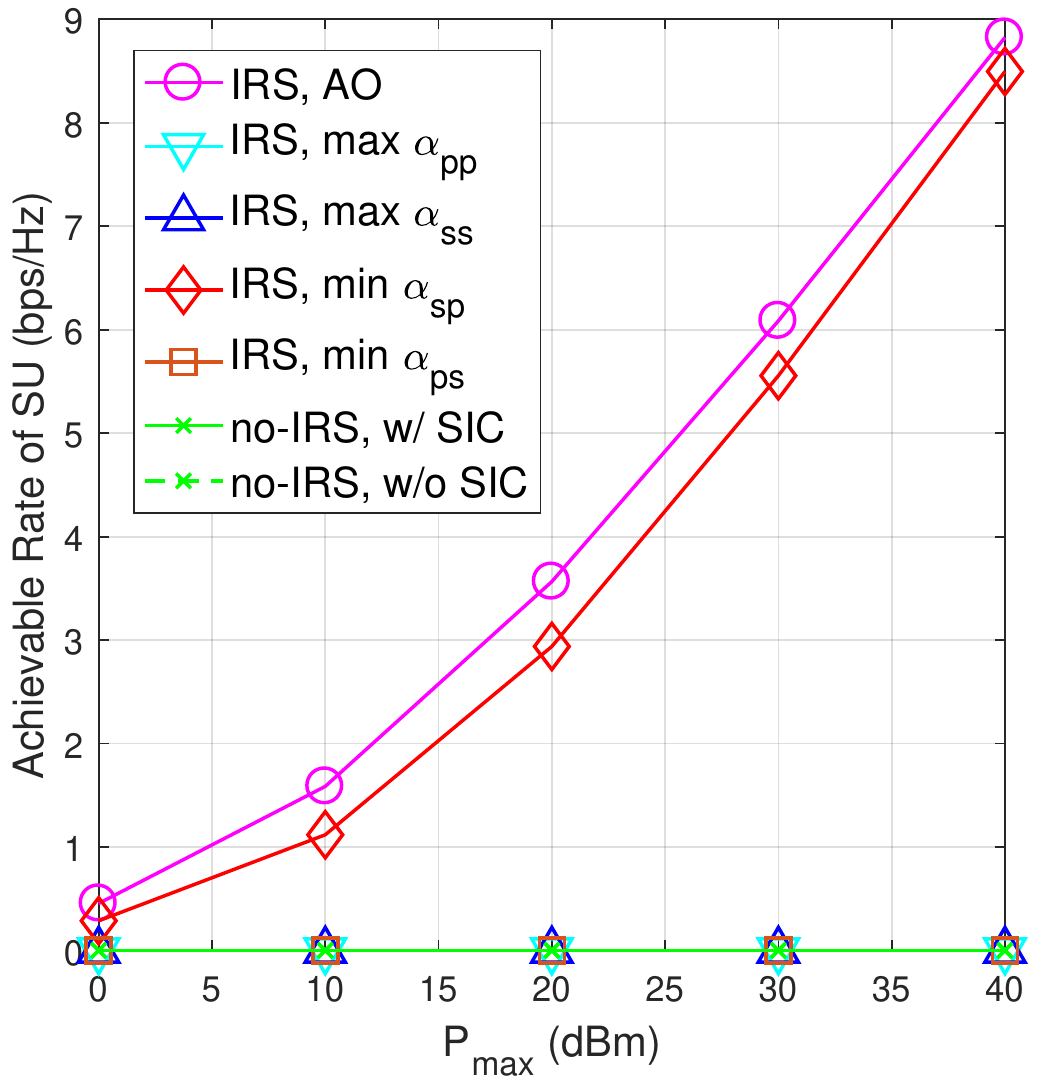}
		\label{fig:rate_setup_c_2}
	}
	\vspace{-2mm}
	\caption{Achievable SU rate for different IRS beamforming designs, with  $N\!=\!60$, $P_0\!=\!20$ dBm, and $\gamma_\text{th}\!=\!20$ dB.}
	\label{SU_Rate_Vs_P}
	\vspace{-5mm}
\end{figure*}
Fig. \ref{SU_Rate_Vs_P} shows the achievable SU rate versus the maximum transmit power of the ST in the three setups. In Setup (1), the SU rate for the cases of (no-IRS, w/ SIC) and (IRS, min $\alpha_{ps}$) remains unchanged when $P_\text{max} \ge 20$ dBm, whereas those for the cases of (IRS, AO) and (IRS, min $\alpha_{sp}$) can sustain increasing with $P_\text{max}$. This is because in the cases of (no-IRS, w/ SIC) and (IRS, min $\alpha_{ps}$), though the inference from the PT(P2) to the SR(S1) is mitigated, $p_s$ is still restricted by the maximum tolerable interference at the PR(P1) and thus the SU rate becomes saturated. While in the cases of (IRS, AO) and (IRS, min $\alpha_{sp}$), the interference from the ST(S2) to the PR(P1) is reduced/minimized by IRS reflection, which thus allows the ST(S2) to increase its transmit power for achieving higher SU rate. One can also observe that all low-complexity designs incur substantial rate loss as compared to the AO based one. This implies that in Setup (1), the benefit of using IRS is manifold, rather than solely enhancing or weakening any single signal/interference link. 

In Setup (2), the case of (IRS, AO) always achieves higher SU rate than that of (no-IRS, w/ SIC). The reason is that using IRS not only cancels the interference from the PT(P1) to the SR(S1), but also enhances the desired signal from the ST(S2) to the SR(S1). It is also observed that the case of (IRS, min $\alpha_{ps}$) achieves nearly the same SU rate as that of (no-IRS, w/ SIC), which implies that when both the PT(P1) and the SR(S1) are nearby the IRS, the low-complexity design minimizing $\alpha_{ps}$ cancels the interference from the PT(P1) to the SR(S1) effectively. However, the other three low-complexity designs are all ineffective for dealing with this strong interference scenario and thus the achievable SU rate is 0. Moreover, note that the SU rate for the case of (IRS, AO) eventually saturates. This is because both the ST(S2) and the PR(P2) are not in the coverage of IRS and thus the interference from the ST(S2) to the PR(P2) cannot be reduced.

Finally, for Setup (3), both no-IRS designs are ineffective, regardless of whether SIC is applied at the SR(S2) or not. This is expected since given the severe interference from the ST(S1) to the PR(P1), the former should keep silent to guarantee the QoS of the latter. This is also the most challenging scenario in conventional CR systems without using IRS. While with our proposed designs (IRS, AO) and (IRS, min $\alpha_{sp}$), such interference is reduced by IRS reflection, thus the SU can access the spectrum and achieve higher rate with increased transmit power. Moreover, similar to no-IRS designs, the other three low-complexity designs are also ineffective because none of them can handle the severe ST-PR interference in this setup.

\vspace{-3mm}
\section{Conclusion}\vspace{-1mm}
In this letter, we studied the SU rate maximization problem via the joint transmit power control and IRS reflect beamforming. We developed an AO-based algorithm to solve the problem efficiently and showed by simulations the effectiveness of employing IRS to improve the SU rate and its advantages in dealing with highly challenging interference scenarios in conventional CR systems without the IRS. 
\begin{appendices}	
	\vspace{-4mm}
	\section{}\vspace{-1mm}
	Consider the function $f(x,y)={|x|^2}/{y}$, $x \in{\mathbb{C}}$, $y\in{\mathbb{R}^{++}}$. Since $f(x,y)$ is jointly convex w.r.t. $(x,y)$, applying the first-order Taylor expansion at $(x_0,y_0)$ yields\vspace{-1mm}
	\begin{equation}\small
	\label{f_x_y}
	f\left( {x,y} \right) \ge {{2{\mathop{\rm Re}\nolimits} \left\{ {x_0^*x} \right\}}}/{{{y_0}}} - {{{{\left| x \right|}^2}}}y/{{{y_0^2}}}.\vspace{-1mm}
	\end{equation}
	Setting $x={\bf{\bar h}}_{ss}^H{\bf{\tilde v}}$, $x_0={\bf{\bar h}}_{ss}^H{\bf{\tilde v_0}}$, $y={{{{{\bf{\tilde v}}}^H}{{\bf{B}}_{ps}}{\bf{\tilde v}}}}$ and $y_0={{{{{\bf{\tilde v}}}_0^H}{{\bf{B}}_{ps}}{\bf{\tilde v_0}}}}$, we have \vspace{-2mm}
	\begin{equation}\vspace{-2mm}
	\footnotesize 
	\label{f}
	f\left({{\bf{A}}_{ss}},{{\bf{B}}_{ps}}, {\bf{\tilde v}} \right) \!\ge\! \frac{{2{\mathop{\rm Re}\nolimits} \left\{ {{\bf{\tilde v}}_0^H{{\bf{A}}_{ss}}{\bf{\tilde v}}} \right\}}}{{{\bf{\tilde v}}_0^H{{\bf{B}}_{ps}}{{{\bf{\tilde v}}}_0}}} \!-\! \frac{{{\bf{\tilde v}}_0^H{{\bf{A}}_{ss}}{{{\bf{\tilde v}}}_0}}}{{{{\left( {{\bf{\tilde v}}_0^H{{\bf{B}}_{ps}}{{{\bf{\tilde v}}}_0}} \right)}^2}}}{{{\bf{\tilde v}}}^H}{\bf{B\tilde v}}.
	\end{equation}
	By applying Lemma 1 in \cite{Palomar}, an upper bound on ${\small\!{{{\bf{\tilde v}}}^H}{\bf{\small {B}\tilde v}}\!}$ is given by \vspace{-0.5mm}
	\begin{equation*}
	\small
	\label{vBv}
	\!{{{\bf{\tilde v}}}^H}{\bf{{B}\tilde v}}\!\! \le\! 2{\mathop{\rm Re}\nolimits} \!\left\{\! {{\bf{\tilde v}}_0^H\left( {\!{{\bf{B}}_{ps}}\! \!\!-\!\! {\lambda _{{{\bf{B}}_{ps}}}}{{\bf{I}}_{N \!+\! 1}}} \!\right)\!{\bf{\tilde v}}} \!\right\} \!+ 2{\lambda _{{{\bf{B}}_{ps}}}}\!\left(\! {N \!\!+\! \!1} \right) \!-\! {\bf{\tilde v}}_0^H{{\bf{B}}_{ps}}{{{\bf{\tilde v}}}_0}.
	\end{equation*}\vspace{-5.5mm}\\
	Substituting it  into (\ref{f}), we obtain (\ref{lemma1}).
	\vspace{-3mm}
	\section{}\vspace{-0.5mm}
	The Lagrangian associated with (P2.3) is expressed as \vspace{-4mm}
	
	{
		\small
		\begingroup
		\addtolength{\jot}{-0.5mm}
		\begin{align}	
		L\left( {{\bf{u}},\lambda ,\left\{ {{\mu _n}} \right\}} \right)\!=\!&\sum\nolimits_n {\left( {2{\mathop{\rm Re}\nolimits} \left\{ {\left( {w_{ss}^*(n)\! +\! \lambda w_{pp}^*(n)} \right){u_n}} \right\} \!-\! {\mu _n}{{\left| {{u_n}} \right|}^2}} \right)} \notag\\
		&+ \sum\nolimits_n {{\mu _n}}+{d_{ss}}+\lambda {d_{pp}} \!-\! \lambda {\gamma _\text{th}}.\label{Lang}
		\end{align}\vspace{-5mm}
		\endgroup
	}\\ 
	Accordingly, the dual function is given by ${g}\left( {\lambda ,\left\{ {{\mu _n}} \right\}} \right)= {\sup _{\bf{u}}}L\left( {{\bf{u}},\lambda ,\tiny\{ {{\mu _n}} \tiny\}} \right)$.
	To make $g\left( {\lambda ,\left\{ {{\mu _n}} \right\}} \right)$ bounded from above, i.e., $g\left( {\lambda ,\left\{ {{\mu _n}} \right\}} \right)\!<\!\infty$, it follows that $\mu_n\!>\!0$, $\forall n$, must hold. Denoting the optimal primal and dual variables by ${\bf{u}}^\star$, $\lambda^\star$ and $\mu_n^\star$, $\forall n$, then ${\bf{u}}^\star$, $\lambda^\star$ and $\mu_n^\star$ should satisfy the Karush-Kuhn-Tucker (KKT) optimality conditions. Using the complementary slackness condition for (\ref{C8}), we have ${\mu _n^\star}\big( {1 - {{\left| {{u_n^\star}} \right|}^2}} \big){\rm{ = 0}}$. 
	Since $\mu_n^\star\! >\!0$ always holds, ${u_n^\star}$ has to satisfy $|{u_n^\star}|=1$. By exploiting the first-order optimality condition to maximize the Lagrangian in (\ref{Lang}), $u_n^\star$ with fixed $\lambda$ and $\mu_n$ is derived as $u_n^\star = {{\left( {{w_{ss}}(n) + \lambda {w_{pp}}(n)} \right)}}/{{{\mu _n}}}, n=1,...,N+1.$
	Since $|u_n^\star|\!\!=\!\!1$,  the proof is thus completed.

	\vspace{-2.5mm}
	\section{}\vspace{-0.2mm}
	Denoting $a( n ) \!\!=\!\! \cos \left( {\angle \left( {{w_{ss}}(n) \!+\! \lambda {w_{pp}}(n)} \right) \!\!-\!\! \angle \left( {{w_{pp}}(n)} \right)} \right)$, ${\gamma _p}$ can be rewritten as ${\gamma _p}\left( \lambda  \right) \!= \!2\Sigma_n\left| {{w_{pp}}(n)} \right|a\left( n \right)\! +\! {d_{pp}}$. Suppose that the values of ${w_{ss}}(n)$ and ${w_{pp}}(n)$ are given by $x_1+jy_1$ and $x_2+jy_2$, respectively, $x_1, x_2,y_1,y_2 \in \mathbb{R}$, then $a(n)$ can be expressed as \vspace{-2mm}
	\begin{equation*}\footnotesize\vspace{-1mm}
	a\left( n \right) = \sqrt {1 - {{{{\left( {{x_2}{y_1} - {x_1}{y_2}} \right)}^2}}}/\big({{{a_1^2} + {a_2^2} + {a_3^2} + {a_4^2}}}}\big),\vspace{-1mm}
	\end{equation*}
	where $a_1\!=\!{{\left( {{x_1}{x_2} + \lambda {x_2}^2} \right)}}$, $a_2\!=\!{\left( {{y_1}{y_2} + \lambda {y_2}^2} \right)}$, $a_3\!=\!{\left( {{x_2}{y_1} + \lambda {x_2}{y_2}} \right)}$ and $a_4\!=\!{\left( {{x_1}{y_2} + \lambda {x_2}{y_2}} \right)}$. Since $\angle \big(w_{ss}(n)\big) \!\!\ne\!\!\angle \big(w_{pp}(n)\big)$, i.e. ${x_2}{y_1} \!\ne\! {x_1}{y_2}$, $\forall \lambda\!>\!0$, $a(n)$ strictly increases as $\lambda$ increases, which thus completes the proof.  \vspace{-2mm}
\end{appendices}

\ifCLASSOPTIONcaptionsoff
\newpage
\fi
\bibliographystyle{IEEEtran}
\bibliography{IEEEabrv,reference}

\end{document}